# A Structural Model of Intuitive Probability

**Jean-Louis Dessalles (dessalles@enst.fr)**
ParisTech-ENST (CNRS UMR 5141), 46 rue Barrault
F-75013 Paris - France

**Abstract**

Though the ability of human beings to deal with probabilities has been put into question, the assessment of rarity is a crucial competence underlying much of human decision-making and is pervasive in spontaneous narrative behaviour. This paper proposes a new model of rarity and randomness assessment, designed to be cognitively plausible. Intuitive randomness is defined as a function of structural complexity. It is thus possible to assign probability to events without being obliged to consider the set of alternatives. The model is tested on Lottery sequences and compared with subjects' preferences.

## Introduction: Cognition and Probability

There seems to be an irreducible divorce between mathematical probability and human abilities to estimate odds. Various studies in psychology have revealed spectacular biases that seem to prove that human beings do not have a genuine consistent competence in probability. For instance, the gambler's fallacy induces individuals to avoid betting on tail when tossing a coin if tails have been drawn several times in a row, as if the coin was expected now to fall on face to restore balance (Terrell 1994). Another error, on which much has been written, is known as the conjunction fallacy (Tversky & Kahneman 1983). It consists in considering more likely that Linda, an anti-war militant female student, be found a few years later to be (a) a bank teller and militant feminist, rather than (b) a bank teller (without further precision). Normally, the latter case should be considered more probable, as it includes the former. A third famous error, called base-rate fallacy, consists in inferring probability from immediate evidence without taking *a priori* rarity into account, as when someone is sure to be ill when a biological test with less that 1% error says so, though that person has 33% chances to be in good health if the illness concerns only 2% of the population (Evans 1990). Moreover, people are over-sensitive to data presented as concrete and close and tend to ignore frequency statements (Nisbett & Ross 1980).

Though it can be argued that such errors are unlikely to show in everyday life and need artificial testing situations to be revealed, they still seem to dismiss any possibility that human cognition involve a true competence in probability.

The problem is not only on the cognitive side. By various aspects, classical mathematical probability theory is unable to provide an acceptable theoretical frame for the corresponding cognitive abilities. Let us mention two reasons for this.

First, there are several cognitive modes of access to probability that are artificially conflated into one single concept through axiomatics and measure theory. One is symmetry: we feel compelled to assign a probability of 1/6 to each side of the dice, just because the dice is symmetrical. Another one is atypicality: if we see a cat which is three standard deviations away in size from our prototype of cat, we consider it improbable. Yet another is provided by frequency estimates in time, in space or in any relevant dimension: if all the fires I witnessed were located kilometres away from my home, I infer a spatial density for such events that makes me consider a fire next block quite improbable. From a cognitive modelling perspective, it is not obvious that these modes of access to probability can be unified the way classical probability does it, *i.e.* using measure theory, distributions and basic axioms.

Second, the foundations of classical probability theory are problematic from a cognitive modelling perspective. Authors like Franck Ramsey and Bruno de Finetti attempted to ground probability into intuition. They independently designed a procedure to found the concept of subjective probability (Gillies 2000). Their idea was to imagine an experiment, in which the subject is asked to bet on the event. The subject does not know in advance whether it is a bet in favour of the event or against it. The relative height of the bet happens to reflect the subject's estimate of the probability. Moreover, such estimates are shown to satisfy Kolmogorov's axioms. The problem is that all classical definitions of probability, including de Finetti and Ramsey's subjective probability, require explicit knowledge of all alternatives. De Finetti's provocative statement, "Probability does not exist", claims that probability is grounded in the observer's ignorance, not in some objective knowledge (de Finetti 1970). Though, the 'subjective probability' framework still requires some objective reference. To assign a probability to a given event, one must know the complete set of alternatives. How can non-omniscient beings assign probability at all?

Let us take an example. Suppose a reader notices an unexpected fact about the novel she is currently reading. All six paragraphs in the current page begin with the character 'C'. The fact may be perceived as unusual by that reader because, obviously, she considers that the page could or should have been otherwise. What is the set of alternatives there? Alternative capital letters, alternatives letters, alternative paragraphs, alternative pages, alternative novels? What does prevent the reader from being overwhelmed with irrelevant details when considering alternatives, such as paper irregularities, scattered tiny ink flecks and the like? After all, the actual page is unique, and thus highly improbable, due to many tiny details that can be discovered in it. Maybe in no other printed page, in the history of humanity, did the four words 'excited', 'deep', 'talked', 'gone' appear one above the other on four adjacent lines. Though, this latter fact would appear exceedingly boring to most of us. And the reason lies in the cognitive perception of probability. The repeated character 'C' at the beginning of all six paragraphs of the page appears more improbable than the vertical arrangement of four particular words. We must explain why.

In what follows, we start from Solomonoff's definition of algorithmic complexity, and we adapt it to cognitive modelling purposes. Then we define unexpectedness and base a new definition of intuitive probability upon it. The prediction that human subjects will find unexpected structures improbable will be explored in a little exploratory experiment. Finally, we will discuss the generality of the notions introduced in this paper.

## A Structural Approach To Probability

In the framework of algorithmic complexity developed by Ray Solomonoff, Andrey Kolmogorov and Gregory Chaitin, the probability of an object is defined in relation to the shortest available descriptions of that object. The epistemological shift is significant. Probability is no longer a measure of the subject's ignorance about the object, as in classical definitions. On the contrary, it depends on the subject's ability to describe the object. More precisely, the algorithmic probability $P_M(x)$ of an object $x$ is given by (Solomonoff 1978):

$$P_M(x) = \alpha \sum_i 2^{-L(s_i)} \quad (1)$$

This definition is relative to a given Turing machine $M$, and $L(s_i)$ is the length of the $i^{th}$ string that, given as input to that machine, outputs a string beginning with $x$. $\alpha$ is a normalisation coefficient that depends on $M$ and on the size of $x$. This probability is well approximated by:

$$P_M(x) \approx 2^{-K(x)} \quad (2)$$

where $K(x)$ is the Kolmogorov complexity of $x$.[1]

The concept of algorithmic probability captures the odds that a given device will spontaneously produce object $x$ ('spontaneously', in the case of a Turing machine, means: when given a random sequence as input). For this definition to be of any use in cognitive modelling, the machine $M$ has to be replaced by some model of the human cognitive ability to analyse objects. In the spirit of the definition of algorithmic complexity, we may say that any realistic computational cognitive model would be appropriate, as the resulting definitions will only differ by the complexity of discrepancies between these models, which is expected to be small.

We choose Michael Leyton's Generative Theory of Shape as reference model (Leyton 2001). According to Leyton, individuals analyse the structure of objects by constructing it as a multi-layered group. At each stage of this construction, a *fiber* group is transferred through a *transfer* group to give a higher-level structure. A simplified version of this theory will allow us to compute de complexity of structures. Consider for instance the following numeric sequence:

10 20 30 40 50 60 70

This sequence does not appear as random to a human mind trained to read numbers. One possible generative description of this sequence is as follows:

| | | | | | | | |
|---|---|---|---|---|---|---|---|
| 1. start with one uninstantiated number: | _ | | | | | | |
| 2. transfer it through translation: | _ | _ | _ | _ | _ | _ | _ |
| 3. the number is a duplicated digit: | __ | __ | __ | __ | __ | __ | __ |
| 4. instantiate the second digit: | _0 | _0 | _0 | _0 | _0 | _0 | _0 |
| 5. dissociate the transfer operator: | _0 | _0 | _0 | _0 | _0 | _0 | _0 |
| 6. instantiate the first digit: | 10 | 20 | 30 | 40 | 50 | 60 | 70 |

Several generative stories of minimal complexity can be proposed for the same structure, mainly by changing the order of the different phases (*e.g.* step 2 could be postponed after step 4). Valid generative sequences maximise transfer and recoverability (Leyton 2001). In step 5, the transfer operator *cop* which performed a mere translation in step 2 becomes a joint operator *+1/cop*, with one first component that increments its operand.

To compute the complexity $C_R$ of the resulting structure, we propose to use the general formula:

$$C_R = C_F + C_T \quad (3)$$

$C_F$ represents the complexity of the fiber group, while $C_T$ is the complexity of the transfer group. Moreover, the complexity of instantiating a number is given by:

$$C_n = \log_2(n+1) \quad (4)$$

It represents the minimal information in bits to locate the number by its rank in the usual order.[2] Some numbers may have a lower complexity. For instance, the number 33333 results from the transfer of the digit 3, and its complexity is $C_3+C_{cop}+C_5$ rather than log(33334) ($C_{cop}$ designates the complexity of the copy operation, and $C_5$ is the complexity of stopping the copy after reaching 5 terms). In our previous example, each step adds complexity: $C_{cop}$ for step 2; $C_{dup}$ for the duplication in step 3; 0 for the instantiation of step 4;[3] $C_{dup} + C_{+1}$ for splitting the transfer operator and instantiating the first component to the incrementing operator in step 5; and lastly $C_1=1$ for step 6. We may consider that $C_{dup}= C_{cop}$ and that $C_{+n}= C_n$. The overall complexity is thus: $C_R = 3C_{cop}+2$ (we ignore that the sequence is bounded to seven elements, otherwise $C_7$ must be added).

---

[1] This is true only for large objects. A better approximation is given by replacing $K(x)$ by $K_M(x)$, which is the shortest input of $M$ that gives $x$ as input.

[2] Definitions for the complexity of numbers may slightly vary, but those differences are irrelevant to our purpose here.

[3] The complexity of 0 is not necessarily 0, especially if it is perceived as following 9, in which case the complexity is rather 3.5.

If we apply formula (2) with $C_R$ as a human estimate of $K(x)$, we obtain a paradoxical result. Simple structures like 33333 turn out to be more probable than more "random" numbers such as, say, 28561, in flagrant contradiction with what human subjects consider to be the case (see below). The logic underlying definitions (1) and (2) is that simple structures are more likely to be produced by any constructive device than complex ones. This assumption does not capture the human perception of randomness and probability.

In studying apparent randomness, Kahneman and Tversky (1972) observed that subjects require random sequences to be irregular. They mention that apparent randomness "is a form of complexity of structure", as "random-appearing sequences are those whose verbal description is longest." This observation is much in accordance with the modern definition of randomness, which is equivalent to the absence of structure (Chaitin 2001). However, it is at odds with the above definition of algorithmic probability: a regular structure is simple to achieve for usual devices, and should be considered more probable. The problem comes from the fact that human beings consider probabilistic outcomes to be produced, not by simple deterministic devices, but by random, and thus complex, devices. Therefore, we are not interested in some objective probability of the existence of an object, but in the probability of its occurrence as output of a complex machine. This is what the notion of unexpectedness captures.

## Unexpectedness

To account for some important aspects of subjective randomness, we introduce the notion of unexpectedness as follows.

$$U(x) = C_{exp}(x) - C_{obs}(x) \quad (5)$$

where $C_{exp}$ is the expected complexity of object $x$ and $C_{obs}$ is its actual, observed complexity. Let us explain the concept with an example.

If one is given a five-digit number, one expects a complexity $C_{exp} = C_{cop} + 5 \times log(10)$, as a typical number results from a generative process such as (i) copy an uninstantiated digit; (ii) instantiate the five digits independently, after the copy, by choosing among ten possibilities. This is probably the complexity that will be assigned to 28561 by most subjects. If subjects observe that the given number happens to be 33333, they are surprised. This comes from the fact that 33333 results from a simpler process, of complexity $log(10) + C_{cop}$, as the instantiation now occurs before the copy. The amount of surprise is given by $U = 4 \times log(10)$. Such a surprise is generally accompanied by a feeling of improbability (Dessalles, *to appear*). For instance, children may become excited and feel compelled to signal the event when their parents' car reaches 33333 on the clock. To capture this phenomenon, we define subjective probability as follows.

$$p(x) = 2^{-U(x)} \quad (6)$$

In our example, we get $p = 10^{-4}$, which corresponds to the mathematical probability of observing a number with five identical digits. We propose that subjective probabilities defined by (6) provide an adequate picture of the way human individuals perceive probability.[4] We designed a small experiment to test the validity of this claim.

## The Lottery Experiment

We performed a few tests on a small set of subjects as a pilot experiment to test our model. Subjects informally approached in a café were offered a free French Lottery bulletin. They just had to choose 2 among 14 possible combinations of six numbers ranging from 1 to 49 to try their luck. Among these combinations, ten were fixed and four were randomly generated anew for each subject. The resulting lists of combinations were randomly shuffled to avoid artefacts due to the order of presentation. Some of the combinations offered to all the subjects, such as [1 2 3 4 5 6], [10 11 12 44 45 46], [10 20 30 31 32 33], [34 35 36 37 38 39], were chosen for their low complexity. Some others were the output of a random generator.

Twenty-six subjects were tested. Though all combinations are obviously equally likely to win, all subjects showed strong avoidance of the two simplest ones, namely [1 2 3 4 5 6] and [34 35 36 37 38 39]. When asked, all subjects expressed their strong feeling that these combinations were virtually impossible. Typical comments were 'If you play that one, it's sure you won't win' or 'It would be stupid to play that one'. There is only $3.10^{-4}$ probability that the 26 subjects would have consistently missed these two combinations if they had chosen randomly. This result is consistent with findings about the avoidance of highly regular structures by lottery players (Savoie & Ladouceur 1995). It is also consistent with our claim that subjects spontaneously estimate probability through a difference of structural complexity. When the actual structure is too simple, the probability is perceived as too low for the event to have any chance to occur. By contrast, when the structure is complex enough, as for [6 17 21 28 37 42], subjects have no problem choosing the combination.

Though this pilot study is by far insufficient to give reliable results, we tried to estimate how far its results complied with the predictions of formula (6). To do so, we wrote a small computer programme in Prolog to assess the complexity of combinations. The programme maintains a "short-term memory", implemented as a list of four items maximum, which contains the last numbers or operations that were processed. The only operations taken into account are +1 and +2 increments. To reproduce the principle of formula (3), elements that are present in short-term memory are considered "free". Two sequences like 3 4 and 3 4 5 6 7 are thus of same complexity, since the last processed number and the +1 operation are always on top of the short-term memory. The programme detects increments, assigns intrinsic complexity to digits, and processes tenths and units independently if no global increment was previously detected. Though various details of the programme are necessarily arbitrary, they are of no consequence on the overall ranking of structures. Table 1 gives the simplest structures of the sample and their complexity as computed by the programme.

---

[4] Note that in general, it is difficult to verify that subjective probabilities, as defined by (6), comply with Kolmogorov axioms, as they concern events that are not incompatible. For instance, the expected event of five non-related digits is given probability 1, but it includes all more particular events such as 33333.

Table 1: Simple lottery structures of the sample.

| Combinations | Complexity |
|---|---|
| 1  2  3  4  5  6 | 3 |
| 34 35 36 37 38 39 | 6 |
| 10 11 12 44 45 46 | 11 |
| 7  8  9 37 38 39 | 12 |
| 8  9 26 27 28 29 | 12 |
| 10 20 30 31 32 33 | 12 |
| 1  2  5  6 15 49 | 14 |
| . . . | . . . |
| 14 24 36 38 42 44 | 26 |

Figure 1 shows the distribution of the combinations chosen by subjects according to their complexity. Note the total absence of low complexity structures (under 7). Though the pattern is globally compatible with our expectations, we may be surprised by the high score of structures with complexity between 9 and 12. It is likely that a few subjects did not pay attention to the kind patterns that appear on table 1. Some among them did not attempt to maximize the intuitive probability, but applied various alternative strategies, such as choosing combinations that are related to their personal life. For instance, one of them declared he chose combination [10 20 30 31 32 33] just because the last number was the area code of his mate's birth location. Another indicated that he chose [8 9 26 27 28 29] because of the two first digits correspond to his birthday.

Obviously, the test cannot properly function with individuals who are aware of the equiprobability of all combinations. When informally testing colleagues or engineering students, we invariably obtained explicit blind strategies ('just take the first two ones') or meta-strategies ('I choose [1 2 3 4 5 6] to maximize my gain expectancy, as nobody will play it'). By contrast, none of the 26 subjects of the sample seems to have applied such mathematically informed strategy.

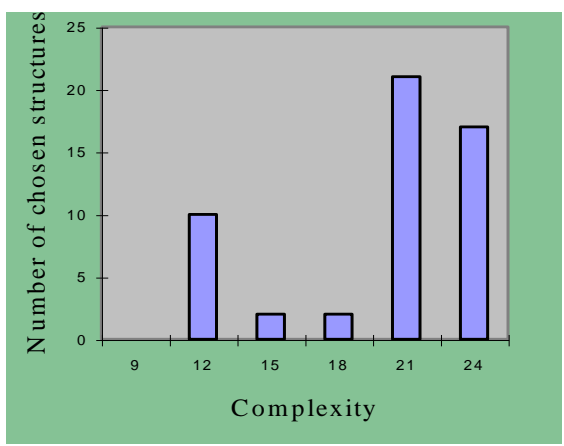

Figure 1: Distribution of chosen structures according to complexity.

This preliminary study provides nothing more than a hint that formula (6) may give a relevant estimate of subjective probabilities. We are planning a new experiment that will correct the shortcomings of this one. In particular, subjects will be made aware of sequential patterns and other conspicuous structures by being shown the Lotto grids as they will be played, together with corresponding combinations in numeric form. Moreover, we plan to investigate the remarkable influence of mathematical instruction on the subjects' choices.

## Discussion

The outcomes of this preliminary study cannot be considered as validated results, and our next experiments with subjects will require significant improvement. However, it demonstrates that the phenomenon is real, that individuals base their judgement concerning probability on various cognitive factors, which include structure. It is not absolute structure that is relevant, but unexpected structure, *i.e.* structure that is less complex than anticipated.

It can be objected that the definition of probability given by formula (6) has no reliable character, as it crucially depends on the observer. For instance, as number like 28561 will appear random to most subjects, while a few of them will consider it as very peculiar, because they recognise the fourth power of 13. What would be a drawback for mathematical theorising (for instance, the French Lottery administration should ignore the structure of drawings) proves fundamental for cognitive modelling. Structure is indeed the only way to understand why subjects are confident in assigning highly diverging values of probability in the absence of any objective reason to do so. It also explains how individuals are able to assign probability to events when the set of alternatives is not defined in advance.

The little experiment on the Lottery reveals what many would consider as a bias. Most individuals, especially if they had no training in mathematical probability, fail to recognise the fact that a combination like [1 2 3 4 5 6] has no less chances to be drawn than [6 17 21 28 37 42] when it comes to making a betting choice. Such 'errors' seem to demonstrate that aspects of human cognition are maladaptive.

In (Dessalles, *to appear*), we show that one of the main occasions on which individuals attend to unexpected states of affairs is to signal them in conversation. For instance, in the middle of a discussion between four people, one participant interrupted the conversation to observe that her three companions were wearing purple shirts. The unexpected decrease of complexity, as the set of purple shirts formed a simpler structure than usually observed within such a group, urged the young woman to make her conversational move. It is not structure *per se*, but rather the fact that she could observe an unexpected transfer (in the Leyton sense), that prompted her utterance.

A possible explanation for this sensitivity to structural complexity drop is that it may be advantageous in normal life, as unexpected structure is quite often correlated to hidden causality. Situations like National Lottery, in which honesty is guaranteed by law, are unlikely to occur in spontaneous settings. If [1 2 3 4 5 6] had been drawn on the very first day when National Lottery was created, most people, including mathematicians, would have been intrigued and would have suspected some hidden causality such as malfunction or malpractice, even if that drawing was as probable as any other, and no less probable than the actual drawing that elicited no suspicious reaction.

More generally, we may wonder to what extent unexpectedness, as defined by (5), is a relevant base for defining subjective probability. One of difficulty in modelling probability is to account for the fact that the same sense of improbability accompanies apparently unrelated phenomena, such as unexpected remarkable structures, atypicality and unexpected proximity. Formula (6) may provide a suitable answer. If we take into account the *complexity of individuation* of situations, then formula (6) offers a unified account of what people would consider improbable. A Lottery drawing like [1 2 3 4 5 6], which is remarkable by its unexpected low structural complexity, is easy to individuate within the set of usual drawings;[5] a one metre tall dog is much easier to individuate that a standard dog; the exact location of a fire that occurred in the immediate vicinity is easier to individuate than the location of remembered fires, which occurred tenths of kilometres away. In each case, the complexity of individuation is measurable. It corresponds to the length of the shortest ideal description that allows distinguishing the situation from any other possible situation. When this complexity is lower than usual, a feeling of improbability ensues, and its amplitude is predicted by formula (6). The same account in terms of complexity drop thus obtains for the various cognitive sources of improbability, while they were merely artificially merged in classical theories.

## Perspectives

As already mentioned, much additional work is needed to validate the model by observing the way individuals behave when dealing with randomness and improbability. On the computer modelling side, we are currently working on algorithms that would offer a more faithful computer implementation of Leyton's principles. The difficulty lies in the programme's ability to detect operators that, when applied, diminish the complexity of the actual structure. Moreover, those operators must be cognitively plausible. The difficulty is that some operations may be easily detected by the computer and not by human subjects, and conversely. For instance, it is not obvious for subjects that the sequence 7 13 18 24 31 39 is remarkable because the gap between numbers is incremented; on the other hand, the mirror structure of a sequence like 2 14 29 35 35 29 14 2 is conspicuous to most subjects, but is not obvious for the computer to detect if it was not specifically programmed to do it. The sequence possesses Leytonian structure, as the pattern 2 14 29 35 is transferred through the symmetry group. The problem is that if the programme is given a variety of operators, we may face a combinatorial explosion when searching for existing transfers in the input structure.

---

[5] It is a general result of algorithmic complexity theory that the number of simple structures of given size tends to be negligible in comparison to complex structures of the same size when that size increases (Li & Vitanyi 1993).

Grounding subjective probability in unexpectedness offers new perspectives on probabilistic reasoning. It obviously provides a new and quantifiable interpretation of some probabilistic biases that were attributed to representativeness (Kahneman & Tversky 1972). It also accounts for the fact that human beings can assign low probability to unique events, without considering the set of alternatives, and it makes non trivial predictions on the influence of parameters like distance or recency on newsworthiness (Dessalles, *in preparation*).

The scope of the notion of unexpectedness, as it has been defined here as a difference of cognitive complexity, goes beyond the sole definition of intuitive probability. Unexpectedness is involved in what makes the interest of conversational narratives (Dessalles, *in preparation*), and in the definition of emergence (Dessalles & Phan 2005). *Cognitive complexity*, understood by analogy with Kolmogorov's definition, as the shortest ideal description, available to human cognition, of perceived objects or scenes, could prove to be an important notion of cognitive modelling.

## Acknowledgments

Special thanks to Hayette Soussou who helped us in the design of the experiment and in carrying out the tests.